# Ovonic switches enable energy-efficient dendrite-like computing

Unhyeon Kang[1,2], Jaesang Lee[1,2], Seungmin Oh[1,3], Hanchan Song[4], Jongkil Park[1], Jaewook Kim[1], Seongsik Park[1], Hyun Jae Jang[1], Sangbum Kim[2], Su-in Yi[4], Suhas Kumar[5*], Suyoun Lee[1,6*]

[1]*Center for Semiconductor Technology, Korea Institute of Science and Technology, Seoul 02792, Korea*
[2]*Department of Materials Science and Engineering, Seoul National University, Seoul 08826, Korea*
[3]*Department of Physics and Astronomy, Seoul National University, Seoul 08826, Korea*
[4]*Department of Electrical and Computer Engineering, Texas A&M University, TX, USA*
[5]*Rain AI, San Francisco, CA, USA*
[6]*Nanoscience and Technology, KIST School, University of Science and Technology, Seoul 02792, Korea*
*Email: S.K. (su1@alumni.stanford.edu), S.L. (slee_eels@kist.re.kr)

*Over the last decade, dendrites within individual biological neurons, which were previously thought to generally perform information pooling and networking, have now been shown to express complex temporal dynamics, Boolean-like logic, arithmetic, signal discrimination, and edge detection for image and sound recognition. Mimicking this rich functional density could offer a powerful primitive for neuromorphic computing, which has sought to replace the aging digital computing paradigms using biological inspirations. Here, using electrically driven Ovonic threshold switching in Sb-Te-doped GeSe, we demonstrate a single two-terminal component capable of self-sustained dynamics and universal Boolean logic, in addition to XOR operations (which is traditionally thought to require a network of active components). We then employ logic-driven dynamics in a single component to detect and estimate the gradients of edges in images, a task that otherwise requires elaborate circuits. A network of Ovonic switches exhibits properties of a half adder and a full adder, in addition to discriminative logic accommodating inhibitory and excitatory signals. We show that this computational primitive is not only seemingly simpler, but also offers many orders of magnitude improved energy efficiency compared to prevailing digital solutions. As such, this work paves the path for potentially emulating dendrites for efficient post-digital neuromorphic computing.*

## 1. Introduction

The early neuron models starting from the 1940s led to a broadly held thought that most of the brain's functions relied on the complex dynamics in the soma, information storage in the synapses, and signal transmission in the axons, while the dendrites of a neuron were thought to primarily perform pooling of information and enabling networking with other neurons.[1-7] Neuromorphic computing, a paradigm that aims at mimicking biological processes to gain the energy efficiency of the brain to replace the performance-saturated digital computing architecture, followed the various early motivations from neuroscience outlined above. For instance, within the past decade, there have been demonstrations of individual electronic components mimicking the various processes within the soma (e.g., information encoded in spikes, bursts, etc.), synapses (storing multi-level information in a nonvolatile fashion), and the axon (moving information without data losses).[8-12] Although neuromorphic computing has closely followed various neuroscientific advances, there has been a large time lag between the two. As an example, emulating the various functions of the soma within a single electronic component, as described by the Hodgkin-Huxley model from the 1950s, was achieved only in 2020.[8] As such, for neuromorphic computing to truly exploit the energy efficiencies of the brain, we need to rapidly incorporate neuroscientific advances into electronic chips.

Over the past few years, neuroscience has shed light on the remarkably complex and critical functions of dendrites, in addition to their known information pooling functions (e.g., efficient fan-out). For instance, as shown by Gidon *et al.*, dendrites within a single typical human L2/3 neuron generate calcium-mediated action potentials, which are capable of independently computing linearly non-separable XOR functions, in addition to employing Boolean-like logic (e.g., AND and OR gates).[13] Further, these dendrites also compute information by balancing inhibitory and excitatory signals. Consistent with these insights, it has also been conjectured that dendrites could calculate gradients in visual images as well as auditory signals, thereby allowing convolution operations, a key process in signal recognition.[14,15] This summary of dendritic functions (Fig. 1), though reductionistic, clearly indicates that dendritic computations are useful and may offer a primitive for functionally dense post-digital neuromorphic computing.

In this work, we sought to combine these dendritic functions (Fig. 1b-f) into a single electronic component and demonstrate that a network of such components could perform more complex functions. These functions involve both analog behaviors as well as digital-like logic. To achieve both sets of functions, we strategically employ Ovonic threshold switching (OTS) in GeSe-based materials. These materials exhibit a purely electric-field-driven volatile electrical switching,[16] making them capable of both volatile instabilities as well as low leakage currents (which is crucial to enable digital-like logic). We specifically engineer Sb-Te-doped GeSe compounds, which, due to amorphous phase stabilization, are particularly suited to exhibit both properties. By biasing at the instabilities within the volatile switching, were demonstrated self-sustained spiking behaviors in an artificial neuron based on OTS.[17]

By constructing simple circuits made of a single active component, along with a few passive capacitors and resistors, we demonstrate all possible Boolean logic gates (specifically, universal logic gates exemplified by a NAND gate). Notably, we demonstrate an XOR gate built using a single active component, a function that usually requires many active components. Using the XOR function as a primitive, we demonstrate a half adder and a full adder as well. We also show how a simple network of two active components can handle one inhibitory and two excitatory signals, similar to the dendrites within the human L2/3 neurons. We then show that a single component, employing its XOR properties, can detect edges in images and also estimate the strength of the edges (via a single-step gradient calculation), which is an important image processing primitive. Finally, we show by experimentally calibrated simulations that the energy efficiency of this solution is many orders of magnitude improved compared to state-of-the-art digital processors.

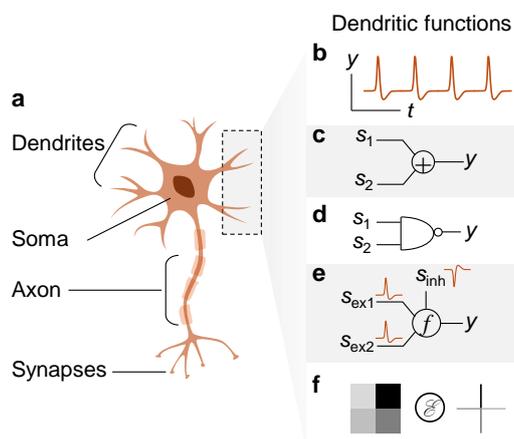

**Figure 1. Overview of the functions of a dendrite. a,** Schematic illustration of a neuron with its general components identified. **b-f** illustrate various functions of a single dendrite in addition to dendritic information pooling: **b,** self-sustained dynamics, **c,** XOR, **d,** universal Boolean logic (illustrated by a NAND gate), **e,** balancing of excitatory signals ($s_{ex}$) and inhibitory signals ($s_{inh}$), and **f,** edge detection and gradient estimation for image processing.

## 2. Results and Discussion

### 2.1 Artificial dendrite based on OTS

To realize both digital logic and analog (i.e., hybrid) functions, we sought an active material with low leakage, low power, high endurance, minimal thermal dissipation, and high speed. Although Mott-insulator-based systems have been used before to demonstrate various neuron-like functions,[18,19] they are unsuitable for hybrid operations since they are primarily driven by Joule heating. Such a process, which incurs large leakage currents, makes them unsuitable to handle Boolean-like operations on zeros and ones (where multiple overlapping zeros, each with a large leakage, could lead to current equivalents to ones). GeSe-based materials fit our requirements in general, which undergo a transition driven purely by electric fields, mediated by metastable metavalent bonds.[20] However, GeSe alone does not offer sufficiently low leakages that can support both logic and analog functions. A key finding of the metavalent-bonds-mediated mechanism was that maintaining an amorphous phase throughout the transition was crucial in minimizing leakage currents in the high-resistance state and overall low energy corresponding to the switching process. In other words, crystallization itself consumes excess energy, in addition to potentially large leakage currents. As such, stabilizing the amorphous phase by doping was key, which is why we chose SbTe-doped GeSe. We show, using transmission electron microscopy (TEM) and X-ray diffraction (XRD), that SbTe-stabilized GeSe has a fully amorphous structure (Fig. 2a and 2b).

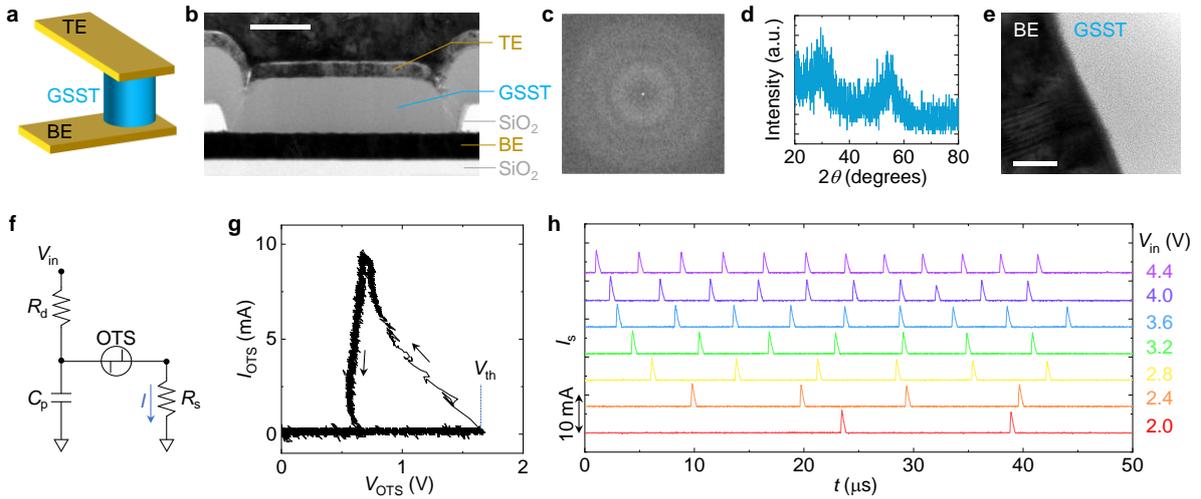

**Figure 2. Artificial neuron device based on Ovonic threshold switch (OTS) and its basic behavior. a,** Schematic illustration of the structure of an OTS device (TE: top electrode, BE: bottom electrode, GSST: Sb-Te-doped GeSe). **b,** Transmission electron microscope (TEM) image of a cross-section of a prototypical OTS device. The scale bar is 100 nm. **c,** Selective area electron diffraction pattern of the chalcogenide thin film obtained using a TEM, exhibiting a largely amorphous structure. **d,** X-ray diffraction (XRD) spectrum of the GSST layer, exhibiting weak to no crystallization. **e,** A magnified TEM image of the cross-section of the interface between GSST and BE, exhibiting a sharp interface. The scale bar is 20 nm. **f,** The circuit used to measure the dynamical properties of the OTS device, which includes an input bias ($V_{in}$), a bias resistor ($R_d$), a capacitor in parallel to the OTS ($C_p$), and a series resistor ($R_s$). **g,** Characteristic dynamical current-voltage (I-V) curve of an OTS device measured using a triangular $V_{in}$ ramp. The threshold voltage ($V_{th}$) is marked. **h,** Self-oscillations (in the form of spike trains) of the OTS device measured as a function of $V_{in}$ ($R_d$ = 9.1 kΩ, $R_s$ = 100 Ω, $C_p$ = 1 nF).

Our OTS device consists of a layer of chalcogenide as an active switching material sandwiched between the top and bottom metallic electrodes in a crosspoint pore structure with TiN acting as the bottom electrode (Fig. 2a-b). By studying electron diffraction patterns in a transmission electron microscope (TEM) as well as X-ray diffraction spectra, we confirm the stabilization of the mostly amorphous chalcogenide layer (Fig. 2c-d). We also ensured that the TiN-chalcogenide interface was sharp and clean since TiN can often lead to reactive or diffusive effects on the interfaces (Fig. 2e). The diameter of the pore ($d_{pore}$) was varied from 200 nm to 20 μm, of which OTS devices with $d_{pore}$ = 6 μm was used in our experiments. To choose the active layer, we first considered the requirements of a dendrite-mimetic OTS (e.g., amorphous phase stabilization) and leveraged our previous study that offered a detailed comparison of various OTS materials and their electrical behaviors.[21] We narrowed our search to two candidates – Sb-Te-doped GeSe (GSST) and Sn-doped GeSe. A comparison among their electrical behaviors (Supporting Information Sec. S1) revealed that GSST offered clearly lower off-state leakage currents (in addition to higher endurance of operation), which was critical for our intended logic applications, which led to our choice of GSST.

To measure the dynamical properties of our OTS device, we constructed a circuit (Fig. 2f), which consisted of a bias resistor, a series resistor, and a series capacitor. We first measured the current-voltage (*I-V*) curve of the OTS device by applying triangular voltage pulses. Unlike typical measurements of *I-V* curves, we report the dynamical behavior of the system (that is, measured without waiting for steady states to be achieved at every applied input bias), which captures not only the overall shape of the behavior, but also its instabilities. The *I-V* curve exhibits a region of negative differential resistance (NDR), where the voltage decreases as the current increases, a fingerprint of neuron-like instabilities.[11,22] NDR is expected to drive the system into neuron-like self-oscillations, the fingerprints for which are also visible in the NDR region of the *I-V* curve (specifically, the non-monotonic fluctuations). To investigate the self-oscillations, we held the input voltage at a constant magnitude and recorded the current through the OTS device with time (Fig. 2h). The self-oscillations appear in the form of periodic spikes, the rate of which depends on the strength of the stimulus ($V_{in}$). Spiking oscillations are also a manifestation of integrate-and-fire behavior, wherein the system accumulates some quantity (e.g., charge) for a certain duration after which it fires a spike. As such, the OTS device exhibits the basic behaviors of an artificial neuron, namely NDR, self-oscillations, integrate-and-fire, and rate coding (encoding of the applied bias in the spiking frequency).

## 2.2 Binary logic gates

Having established basic neuron-like behaviors in our OTS devices, we next sought to demonstrate logic operations. To achieve various Boolean logic functions, we made simple modifications to the measurement circuit Fig. 2g by using input biases, static power supplies, and passive electrical components. The basic idea in these circuits is to arrange to enable the voltage drop across the OTS to exceed $V_{th}$ only for the cases satisfying the truth table corresponding to each Boolean logic operator. For example, the AND gate is implemented by combining one OTS, one capacitor, and three resistors (Fig. 3a-b). Two resistors ($R_1$ and $R_2$) act as the input resistance to the inputs ($s_1$ and $s_2$). In this configuration, the voltage across the OTS can be high only if both $s_1$ and $s_2$ are high, satisfying the condition for AND operation. Following a similar process, if a diode is added to each input terminal, the OR gate can be implemented (Fig. 3c-d). As an aside, the similarity between the circuits for the AND gate and the OR gate is intriguing in that they enable the implementation of a convertible Boolean logic circuit if the diode is replaced by a memristor (the conductance of which can be made diode-like or a linear resistor-like depending on its programmed state).[23-26] Such a convertible logic circuit may offer a primitive for associative learning, which also often requires convertible Boolean logic hardware.[27]

In addition, we also implemented NOR and NAND operators (Fig. 3e-h), which entail inversion in addition to AND and OR operations, which we implemented by simply using an additional power supply. The difference in the circuit compositions between the NAND and NOR operators is the same as the difference between those of the AND and OR operators: the presence of diodes. This consistency seems to imply that any complex Boolean logic operations might be implemented by using an OTS, diode, additional power supplies, and a few passive electrical components. Notably, NAND and NOR operations can be used as the basis to construct any Boolean logic gate, making them universal gates.

While all the above operators are linearly separable, operations such as XOR are non-separable (that is, in a two-dimensional plot of the two inputs and one output of an XOR operator, the output values cannot be separated by a single line). XOR is also an important operation in biological systems, which has been observed in neuronal networks.[13,28] For decades, XOR functions were considered impossible to achieve within a single component (either biological or electronic), as it was traditionally believed that this type of computation required at least two layers of processing and additional summing junctions.[29] Inspired by Gidon et al.'s observation of the linearly non-separable XOR function being expressed by individual dendrites,[13] we implemented the XOR operation using a single OTS device (Fig. 3i-j). In this circuit, the OTS device is an ambipolar switch and is turned on only if inputs $s_1$ and $s_2$ are different from each other. In addition, this circuit can also be utilized as the NOT or INVERT operators while keeping one input fixed at high.

To demonstrate the expandability of the Boolean logic primitives discussed above, by combining some logic operators, we implemented a half-adder circuit by combining an AND and an XOR operators (Fig. 3k-l). In a half-adder, the sum and the carry essentially represent the XOR and AND operators, thereby requiring the combination of the circuits corresponding to these two operations. Furthermore, using two XOR operators, two AND operators, and an OR operator, we also constructed a full adder circuit *in silico*, (Supporting Information Fig. S2). These demonstrations clearly show that logic operators based on OTS devices provide versatile building blocks for post-silicon computers.

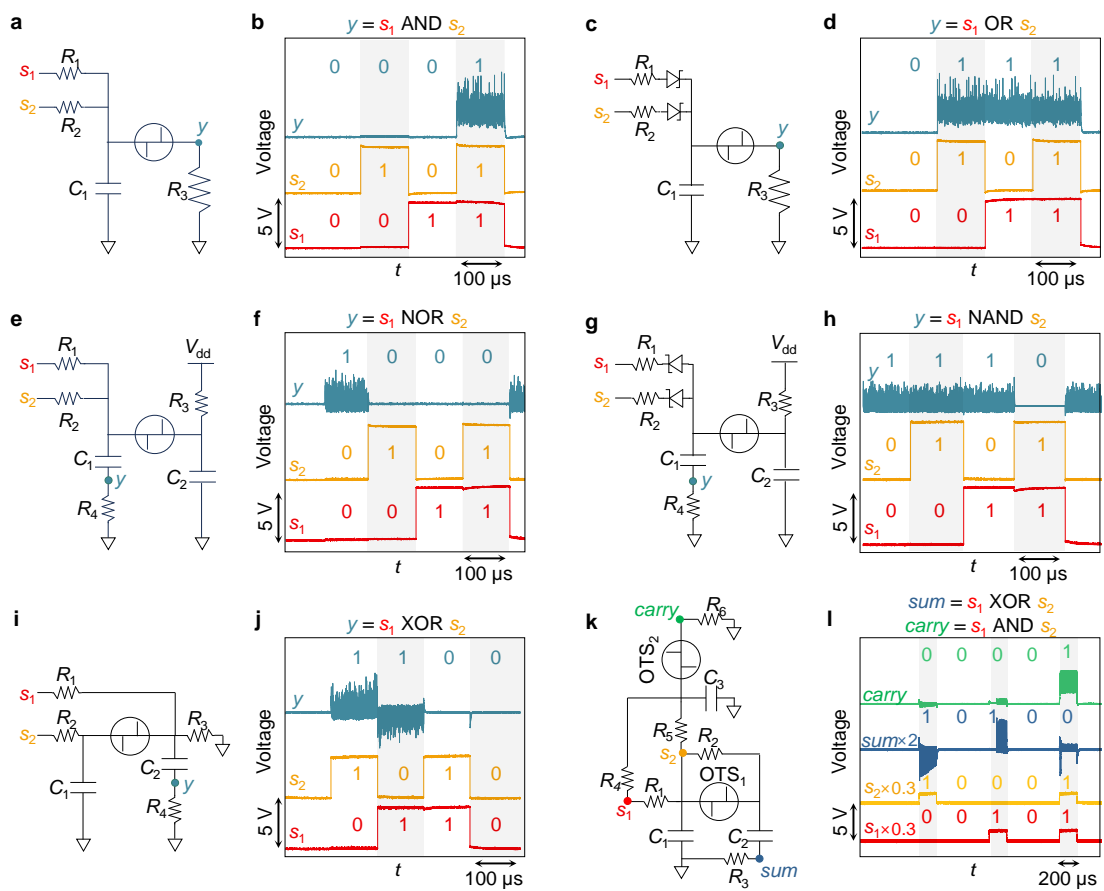

**Figure 3. Five Boolean logic operators and a half adder based on OTS.** The panels present the circuit and the corresponding voltage time-series waveforms. **a-b,** AND gate, with $R_1 = R_2 = 0.9$ kΩ, $R_3 = 5$ kΩ, $C_1 = 100$ pF. **c-d,** OR gate, with $R_1 = R_2 = 0.9$ kΩ, $R_3 = 5$ kΩ, $C_1 = 100$ pF, diode model was1N4744A (Onsemi). **e-f,** NOR gate, with $R_1 = R_2 = R_3 = 0.9$ kΩ, $R_4 = 5$ kΩ, $C_1 = C_2 = 100$ pF, $V_{dd} = 5$ V. **g-h,** NAND gate, with $R_1 = R_2 = R_3 = 0.9$ kΩ, $R_4 = 5$ kΩ, $C_1 = C_2 = 100$ pF, diode model was 1N4744A, $V_{dd} = 5$ V. **i-j,** XOR gate, with $R_1 = R_2 = 1$ kΩ, $R_3 = 50$ kΩ, $R_4 = 10$ kΩ, $C_1 = 1$ nF, $C_2 = 100$ pF. **k-l,** Half Adder, with $R_1 = R_2 = 3$ kΩ, $R_3 = R_4 = R_5 = 1$ kΩ, $R_6 = 200$ Ω, $C_1 = 1$ nF, $C_2 = 100$ pF, $C_3 = 500$ pF.

## 2.3 Mimicking the Ca-mediated dendritic action potential (dCaAP) of pyramidal neurons

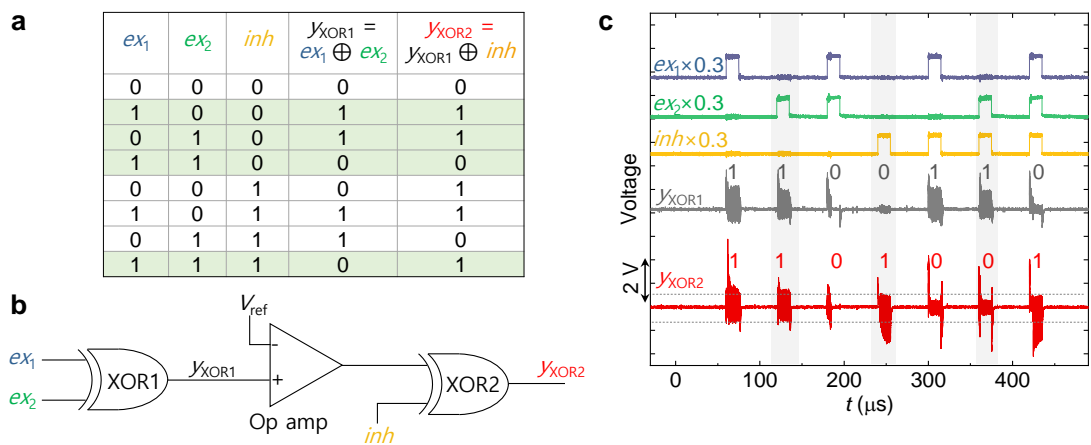

**Figure 4. Mimicking the Ca-mediated dendritic action potential (dCaAP) of human pyramidal neurons. a,** Truth table for a two-stage XOR operator, with two excitatory ($ex_1$, $ex_2$) and one inhibitory ($inh$) inputs. Green-shaded parts were demonstrated in biological pyramidal neurons in the L2/3 region of the human brain.[13] **b,** Circuit for implementing two-stage XOR operation. **c,** Resulting waveforms of the circuit displayed in **b**. Op-amp TLV3602 (Texas Instrument) was used for signal amplification.

As mentioned above, the calcium-mediated dendritic action potential (dCaAP) in the layer 2 and 3 (L2/3) pyramidal neurons of the human brain was reported to exhibit the XOR operation, which we were able to mimic using an OTS device.[13] In addition, it was also shown that the dCaAP is generated when an inhibitory pathway is activated in addition to two excitatory pathways. Such functional dendrites, which can discriminatively combine inhibitory and excitatory signals, are key to not only biological discriminative learning, but are also a keenly studied primitive that may enable highly efficient artificial intelligence (AI) systems via simplified neural network architectures.[30,31] Since our pursuit of OTS devices is motivated by biomimetic computing, we sought to mimic the same truth table that was discovered (via *ex vivo* biological measurements) to be emulated by the L2/3 dendrite-like discriminatory logic (Fig. 4a).[13] The circuit for discriminative dendritic computing was composed of two-stage XOR operators and two operational amplifiers (op-amps) (Fig. 4b and detailed in Supporting Information Fig. S3a), where the op-amps act as comparators. The output of the first XOR stage ($y_{XOR1}$) is amplified to be used as an input to the second XOR stage (the output of which is $y_{XOR2}$). The result (Fig. 4c) presents the output waveforms for eight combinations of the inputs – two excitatory ($ex_1$, $ex_2$) and one inhibitory ($inh$)). The outputs match the truth table in Fig. 4a. In a few cases, for instance, ($ex_1$, $ex_2$, $inh$, $y_{XOR1}$, $y_{XOR2}$) = (1, 1, 0, 0, 0), (1, 0, 1, 1, 0), and (0, 1, 1, 1, 0), the output ($y_{XOR2}$) deviates from ideal values due to distortions, which is attributed to both the nonidealities in the op-amps and the imperfect impedance matches between the OTS devices and the op-amps. We examined the behavior of the circuit by replacing those op-amps with software-emulated op-amps. In the software-assisted experiments, the outputs were closer to ideal outputs (see Supporting Information Fig. S3). Therefore, dendrite-mimicking Boolean logic circuits (and neuronal electronic circuits in general) have the potential not only as fundamental components for brain-like computing, but may also provide an experimental basis to emulate and test neuroscientific theories.

**2.4 Edge Detection**

As the role of dendrites in performing individual but non-trivial computational operations was discovered, the practical utility of such operations also started to emerge. For example, it was recently shown that dendrites are capable of calculating gradients in auditory signals, which explained the previously discovered ability of neurons in calculating gradients (i.e., estimating edge strengths) in images.[14,15] Needless to say, we found edge estimation to be both of practical utility as well as a natural extension of the logic operators demonstrated by our OTS devices. As a benchmark to perform this test, we selected a photograph (512 × 512 pixels) of a mandrill (Fig. 5a).[32-36] For detecting the edges, the original color image was converted into a black-and-white (BW) image (Fig. 5b, see Methods section for details). To implement the edge detection, the BW image was shifted in the horizontal and the vertical directions by one pixel, respectively, rendering the horizontal and vertical opponent images, as illustrated in Fig. 5c. Then, these images were converted into two pairs of one-dimensional (1d) vector arrays ($x_0$, $x_h$ or $x_v$), where the subscript 0, h and v) represent the original, the horizontal and the vertical) opponent images, respectively. These vectors (represented by pulse trains) were experimentally fed as two inputs to the XOR OTS circuit, which generated spikes only when the two inputs were opposite to each other (Fig. 5d), which indicated a pixel on an edge (that is, the two pixels next to each other have measurably different intensities). After detecting the edges in each direction, the resultant 1d arrays were converted back to images, and the two horizontal and vertical edges were combined by pixel-by-pixel OR operations, resulting in the final edge image displayed in Fig. 5e (see Supporting Information Fig. S4, Fig. S5, and Movie S1 for more details). We also conducted edge-detection by using XOR operations defined in software (run on a standard central processing unit) as a reference (Fig. 5f). We found zero errors between the experimental result and the reference. An additional test with another image (a photograph of a few zebras) is presented in Fig. S6 in the Supporting Information.

In addition to detecting edges, we leveraged the analog nature of the rate-coding scheme in an OTS device to enable an edge-detecting OTS system to also estimate the gradient of edges. To this end, we tested edge detection between four distinct grayscale levels (Fig. 5g and Supporting Information Fig.

S7). As expected, the firing rate of the XOR output was observed to be linearly proportional to the contrast difference ($\Delta C$) with a slope of ~ 39.4 kHz per contrast difference and the lower bound of ~ 70 contrast difference for sensing (Fig. 5h). Therefore, two inherent behaviors of a single OTS device – rate coding and XOR operation – can be used to both detect as well as quantify the gradient in edges, not too dissimilar to how dendrites are thought to detect gradients. This mechanism fundamentally differs from the approach employed by conventional CPU-based XOR or edge detection methods.

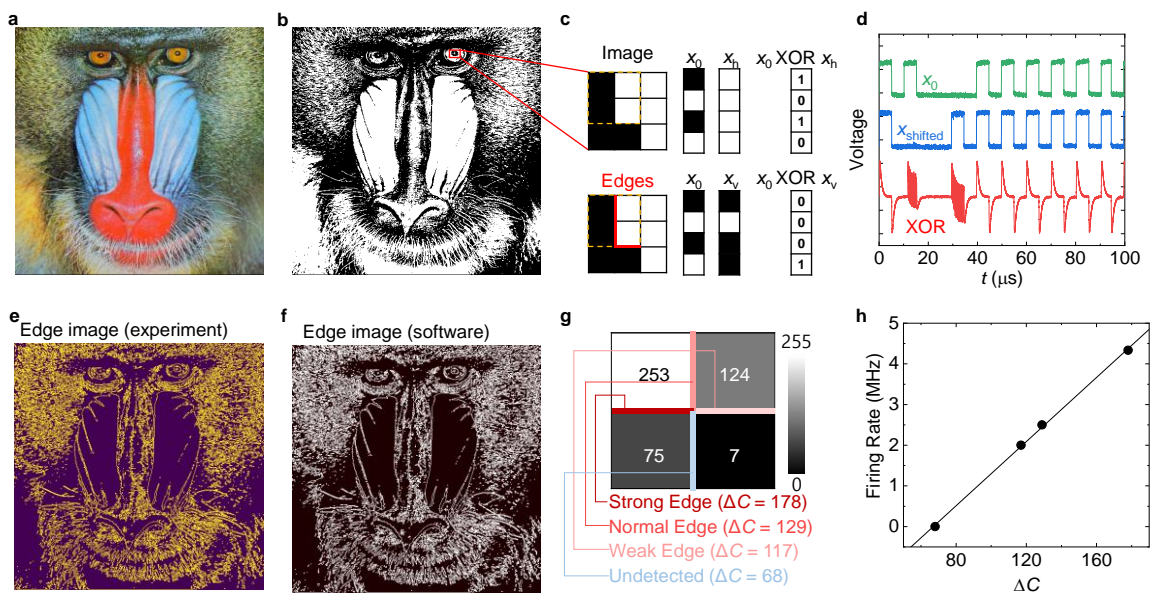

**Figure 5. Experimental demonstration of image edge detection using the XOR neuron. a,** Original image of a mandrill (512×512, 24-bit color-depth). **b,** Black & white (BW) converted image for edge detection (see the Methods for details). **c,** Expansion of a part of the BW image, the horizontally-shifted image ($x_h$), the vertically-shifted image ($x_v$), and the results of an XOR operation on them are illustrated. The red lines indicate the edges detected. **d,** Representative experimental waveforms of two inputs (original and shifted) and the output of the XOR OTS circuit. **e,** Final edge image obtained by the XOR OTS circuit. **f,** Edge image obtained by software calculation, for comparison. The difference between experimental and software calculations was zero. **g,** Four arbitrary gray-scale pixels ranging from 0 to 255, with each pixel value displayed at the center (253, 124, 75, 7). Four strengths of edges (strong, normal, weak, and undetected) are represented, with the contrast differences ($\Delta C$) between pixels indicated. In our edge gradient estimation scheme, as the number of spikes increases, the edge strength becomes stronger. **h,** The firing rate of the XOR operator as a function of $\Delta C$. The solid line is a linear fit.

To assess the energy efficiency of our OTS-based XOR operator, we estimated the energy consumption of the device and conducted a comparison study with those of high-performance digital processors. As a first step, we estimated the experimental energy of our XOR circuit at around 470 pJ/spike (resulting when an edge is detected). To detect edges in a 512×512 image, with shifting of pixels in the horizontal and vertical directions (that is, incorporating a factor of 2), we estimate an energy consumption of roughly 245 µJ. We then estimated the energy consumption for an identical task on several commercial processors – Tesla K20 (GPU, Nvidia), V100 (GPU, Nvidia), H100 (GPU, Nvidia), Xeon E5-2650 (CPU, Intel) (summarized in Table 1). In many of these digital processors, edge-detection is performed using other techniques. For instance, a 3×3 Sobel filter is often used, which we also found to be the most energy efficient (detailed in Supporting Information Table S1) to implement on these processors (thus, we report the energy associated with this technique).[37] These estimates show that the energy consumption of our XOR operator (based on experimental data) is comparable to that of a V100 GPU from 2017 (354 µJ, manufactured at the 12 nm node). To estimate the projected performance of the OTS edge detector at scaled node sizes, we employed the finding in our prior work that showed the energy consumption $\propto d^{1.6}$ in an OTS-neuron device with nearly the same structure, where $d$ is the size of the OTS.[17] This scaling rule is consistent with reliable foundry scaling rules that show a trend of $P = f^n$, where $P$ and $f$ are the power consumption and the feature size, respectively, with $n$ ranging from 1.6 to 2.1 depending on $f$ (detailed in Supporting Information Fig. S9). Using these rules, we project an energy consumption of 3.2 nJ. This energy is over 10,000× lower compared to an Nvidia H100 GPU manufactured at the 4 nm node. We acknowledge the severe overheads in running a relatively general-purpose GPU (and one meant to process larger images) on a specific and small dataset, whereas our solution is bespoke and our estimates do not account for the architectural overheads and data movements. Despite these limitations in our estimates, the fact that we project more than an order-of-magnitude improvement relative to a processor manufactured at a far more advanced node, suggests a potential competitive advantage that cannot be ignored.

|  | K20[38,39] (Nvidia, 2012) | V100[40] (Nvidia, 2017) | H100[41] (Nvidia, 2022) | Xeon E5-2650[38,39] (Intel, 2012) | OTS-XOR ($d = 6$ μm, exp.) | OTS-XOR ($d = 16$ nm, cal.) |
|---|---|---|---|---|---|---|
| Energy per operation (pJ) | 290 | 75 | 20 | 2071 | 467 | 0.356 |
| Edge-detection method | Sobel filter (3×3) | | | | XOR | |
| Number of operations (512×512 pixels) | 4,718,592 (512×512×9×2) | | | | 524,288 (512×512×2) | |
| Total energy consumed (μJ) | 1368 | 354 | 94 | 9772 | 245 | 0.0032 |

**Table I. Energy efficiency of the XOR neuron compared to other technologies.** Energy consumption is estimated for edge detection task on a 512×512 image using different processors: K20 (Nvidia, 2012), V100 (Nvidia, 2017), H100 (Nvidia, 2022), Xeon E5-2650 (Intel, 2012), and our OTS-XOR circuit.

## 3. Conclusion

In conclusion, dendritic functions are less studied compared to the functions of the other parts of a neuron (such as the soma and the synapses). Dendritic functions could offer useful computational primitives, which may improve neuromorphic computing towards the long-sought functional density that is needed to replace the aging digital computing paradigm. Using Ovonic threshold switching materials, we demonstrate a set of dendrite-inspired functions within a single electronic component – universal Boolean logic, XOR function, and edge detection. In a network of such Ovonic switches, we demonstrate half adder, full adder, and dendrite-like discrimination between inhibitory and excitatory signals. We also show that the projected performance of our system is remarkably better compared to state-of-the-art digital processors, in addition to needing fewer active components. The findings of this work broadly align neuromorphic computing with the idea of a smart neuron, which is a neuroscientific principle that neurons perform many more operations than previously modeled many decades ago.

**Methods**

**Fabrication and characterization of OTS and neural operator circuits**

An OTS device consists of an amorphous chalcogenide as an active switching material sandwiched between the top and bottom metallic electrodes. Fig. 2c shows the cross-sectional TEM image of an OTS device used in this work. It has a pore-type structure for controlling the current passing through the device. We have used two different kinds of OTS devices with the structure of Pt/Sn-doped GeSe (SGS)/TiN or TiN/$Sb_2Te_3$-doped GeSe (GSST)/TiN. The composition of SGS and GSST films have been investigated by using Auger Electron Spectroscopy (AES) to be $Ge_{0.34}Sn_{0.18}Se_{0.48}$ and $Ge_{0.4}Sb_{0.17}Se_{0.24}Te_{0.19}$, respectively. As the second device structure showed enhanced cycling endurance, it has been used for the edge-detection task.

The characteristics of OTS devices and neural operator circuits were examined using a two-channel arbitrary function generator (AFG, Tektronix AFG-3102), a SourceMeter (SMU, Keithley 2635B), and a multi-channel oscilloscope (Tektronix DPO-5104). In the experimental demonstrations of the following neural operators, off-the-shelf electrical components have been used (Zener diode: 1N4744A/Jinan Gude electronic device co. ltd, MOSFET (n-type): LND150N3-G/Microchip Inc.).

**Edge-detection of images using the XOR neuron device**

The edge detection task of images (Fig. 5 in the main text and Fig. S6 in the Supporting Information) was conducted by the procedure described in the following. First, an image was converted to the grayscale by applying a weighted sum of RGB values (0.299 for red, 0.587 for green, 0.114 for blue) to reflect human perception, following the industrial standards by PAL and NTSC (i.e., the Y'UV and Y'IQ models, respectively).[26] Next, by applying the threshold to the grayscale image, it turned into a binary (black and white) image. The binarized image was scanned pixel-by-pixel and concatenated into a 1D array. This pre-processing was conducted using a code written in Python. Two 1D arrays representing the original and opponent (shifted by one pixel) images were fed to two arbitrary function generators (ADP3450, Digilent Inc.), which generated two input pulse trains synchronized by the same trigger signal. The width and period of a pulse were set to 5 and 10 μs, respectively, which represented a pixel in the image. Those input signals were sent to the XOR neuron device. The output of the XOR neuron device was recorded by using an eight-channel oscilloscope (MSO58, Tektronix Inc.). The output spikes were counted during each clock period, which was compared with a threshold releasing a train composed of ones and zeros. Then, the resultant 1D bit train was reshaped into the original-resolution-image-size matrix.

**Acknowledgments**

This work was supported by the Korea Institute of Science and Technology (KIST) through 2E33560 and by the National Research Council of Science & Technology (NST) grant by the Korea government (MSIT) (No. GTL24041-000). In addition, the authors thank the Signal and Image Processing Institute (SIPI), the University of Southern California, and Cuyahoga, the photographer, for providing the photos of the Mandrill and the zebra, respectively. S.Y., H.S. and S.Kumar acknowledge the Laboratory Directed Research and Development program at Sandia National Laboratories, a multimission laboratory operated for the US Department of Energy (DOE)'s National Nuclear Security

**Author contributions**

S.L. and S.Kumar designed and conceived the experiments. U.K. and J.L. fabricated the OTS devices and performed experiments demonstrating the functions of the Boolean logic operators. U.K. and S.Kumar designed and performed the experiment of mimicking the behavior of the dCaAP. U.K. and S.O. performed the edge-detection task of images and composed a Python code for comparison. J.P., J.K., and S.Kim contributed to the analysis of the spike waveforms of the Boolean logic operators. S.P. and H.J. contributed to the analysis of the spike waveforms mimicking the behavior of dAP. H.S. and S.Y. calculated the energy consumption of digital processors and the OTS-XOR operator. All authors discussed the data and participated in revising the manuscript.

**Declaration of interests**

The authors declare no competing interest.